\documentclass[proceedings]{JHEP}
\conference{Nonperturbative Quantum Effects 2000}
\title{Integrable Quantum Field Theories with Unstable Particles}
\author{J. Luis Miramontes\thanks{Based on work done in collaboration with 
O.A.~Castro-Alvaredo, C.R.~Fern\'andez-Pousa, A.~Fring and C.~Korff.}\\
Departamento de F\'\i sica de Part\'\i culas,\\
Facultad de F\'\i sica\\
Universidad de Santiago de Compostela\\
E-15706 Santiago de Compostela, Spain\\
E-mail: \email{miramont@fpaxp1.usc.es}}
\abstract{The structure of a new family of factorised $S$-matrix theories with
resonance poles is reviewed. They are conjectured to correspond to the
Homogeneous sine-Gordon theories associated with simply laced compact Lie
groups. Two of their more remarkable properties are, first,
that some of the resonance poles can be traced to the presence of unstable
particles in the spectrum, and, second, that they involve several
independent mass scales. The conjectured relationship with the simply
laced HSG theories has been checked by means of the Thermodynamic Bethe
ansatz (TBA) and, more recently, through the explicit calculation of the
Form Factors. The main results of the TBA analysis are summarized.
}
\def\rf#1{(\ref{eq:#1})}
\def\lab#1{\label{eq:#1}}
\def\nn{\nonumber \\}
\def\bea{\begin{eqnarray}}
\def\ena{\end{eqnarray}}
\def\be{\begin{equation}}
\def\ee{\end{equation}}
\def\JPA#1#2#3{{\sl J. Physics} {\bf A #1} (#2) #3}
\def\JSM#1#2#3{{\sl J. Soviet Math.} {\bf #1} (#2) #3}
\def\EPJC#1#2#3{{\sl Eur. Phys. J.} {\bf C #1} (#2) #3}
\begin{document}
The construction of solvable models that capture realistic
properties of quantum particles is one of the classical motivations for
the study of  two-dimensional integrable quantum field theories. However,
despite the  fact that almost all known particles are
unstable~\cite{Data}, the vast majority of the integrable models
considered so far lack the feature of including unstable particles. The
aim of this talk is to present a new family of factorised $S$-matrix
theories~\cite{DECAY} conjectured to provide the exact solution for the
Homogeneous sine-Gordon models (HSG)~\cite{MASS,PARA} corresponding to
simply-laced Lie groups. The semi-classical spectrum of these models
consists of a finite number of solitonic particles attached to the
positive roots of a Lie algebra, but only those associated with the simple
roots turn out to correspond to stable particles in the quantum theory. 

Although it is very  difficult to distinguish physically
between an unstable particle of long lifetime and a stable particle
({\it e.g.\/}, the proton),  axiomatic $S$-matrix theory makes a big
distinction between them. The reason is that it is based on asymptotic
states that exist for arbitrarily long time and, hence, can only contain
stable particles. In contrast, the basic property of a long-lived unstable
particle, and the one by means of which it is usually observed, it that it
corresponds to a `resonance' in interactions among the stable particles.
Therefore, if two stable particles scatter at centre-of-mass energy
$\sqrt{s}$ close to the mass of an unstable state with appropriate quantum
numbers, the corresponding $S$-matrix amplitude is expected to exhibit a
complex pole at $s_R= (M -i\Gamma/2)^2 $ in the second Riemann sheet. The
position of the pole is given by the mass,
$M$, and decay width, $\Gamma$, of the resonance, whose inverse is the
lifetime of the unstable particle: $\tau= \hbar/ \Gamma$. If the lifetime
is long or, equivalently, if $\Gamma \ll M$, the form of the pole is given
by the Breit-Wigner resonance formula~\cite{BW}:
\be
S \simeq 1- i{2M\Gamma \over s-M^2 +iM\Gamma} \>.
\lab{BW}
\ee
In a generic theory, unstable particles can also induce other
singularities like complex thresholds and cuts. The simplest would
correspond to the production of the unstable particle together with one
stable particle. However, since the unstable particle decays, this kind
of process would lead to particle production and, hence, it has to be
absent if the theory is integrable. Therefore, we will assume that
resonance poles are the only trace of unstable particles in an
integrable quantum field theory.

There is a different HSG theory for each choice of a compact simple Lie
group $G$ (with Lie algebra $g$) and a positive integer~$k$. It describes
an  integrable massive perturbation of the conformal field theory (CFT)
associated with the coset
$G_k/U(1)^{\times r_g}$, where $r_g={\rm rank\>}G$, or,
equivalently,  of the theory 
$G$-parafermions at level-$k$~\cite{PARA}. When $G$ is simply laced, the central charge
of the CFT and the conformal dimension of the perturbation are given by
\bea
&&c_{\rm CFT}= {k\>{\rm dim\/} G\over k + h_g}- r_g = {k-1\over k+
h_g} h_g r_g
\>, \nn
&&
\Delta=  \overline{\Delta}= {h_g \over k+ h_g}< 1\>,
\lab{CFT}
\ena
where $h_g$ is the Coxeter number of $G$. The defining action of the HSG
models is
\bea
&&S_{\rm HSG}= {1\over \beta^2} \biggl\{ S_{\rm GWZW}[h] 
\nn
&&\qquad + {m^2\over \pi}\> \int d^2 x \>\langle 
\Lambda_+,  h^{\dagger} \Lambda_-  h\rangle \biggl\}\>,
\lab{Act}
\ena
whose equations of motion are non-abelian affine Toda
equations~\cite{MASS}. In $S_{\rm HSG}$, $h=h(t,x)$ is a bosonic field
taking values in $G$ and $S_{\rm GWZW}[h]$ is the gauged Wess-Zumino-Witten
action corresponding to the coset $G_k/U(1)^{\times r_g}$. The parameters
$m$ and $\beta^2 = 1/k + O(1/k^2)$ are the bare mass scale and the coupling
constant, whose quantization is required in order to make sense of the
WZW term. $\Lambda_+$ and $\Lambda_- =
i\vec{\lambda}_\pm\cdot \vec{H}$, are two arbitrary constant elements in
the fundamental Weyl chamber of the Cartan subalgebra of $g$ associated
with the maximal torus $U(1)^{\times r_g}$. The HSG theories are quantum
integrable for any choice of $G$ and for any value of
$\Lambda_\pm$~\cite{PARA}, which, in the quantum theory, imply the
existence of $2r_g-1$ different mass scales. The HSG theories are not parity
invariant unless $\Lambda_+$ and
$\Lambda_-$ are chosen to be parallel. 

The semiclassical spectrum of the $G_k$--HSG theory consists of towers of
$k-1$ soliton particles attached to each positive root $\vec{\alpha}$  of
$g$ with masses~\cite{HSGSOL}
\be 
M_{\vec{\alpha}} [n] = {k\over \pi}\> m_{\vec{\alpha}}\> \sin\Bigl({\pi 
n\over k}\Bigr)\>,
\quad n= 1,\ldots, k-1\>.
\lab{MassS}
\ee
For a fixed root $\vec{\alpha}$, this spectrum is 
identical to that one of the minimal $A_{k-1}$ $S$-matrix theory, which
corresponds to the $SU(2)_k$--HSG model. However, the overall mass
scale 
\be
m_{\vec{\alpha}} = 2 m \sqrt{(\vec{\alpha}\cdot \vec{\lambda}_+) 
(\vec{\alpha}
\cdot \vec{\lambda}_-)}
\lab{MassF}
\ee
is different for each tower of particles. Consider now the unique
decomposition of a positive root as a linear
combination of simple roots: $\vec{\alpha} =
\sum_{ i=1}^{r_g} p_i\> {\vec{\alpha}}_i$. Then, eq.~\rf{MassS} satisfies
\be
M_{\vec{\alpha}} [n]\geq \sum_{i=1}^{r_g} M_{\vec{\alpha}_i} (np_i) \>,
\lab{Decay}
\ee
which indicates that the soliton particle $(\vec{\alpha},n)$ is unstable and
decays into particles associated with the simple roots~\cite{DECAY}.
Moreover, for any three roots $\vec{\alpha}$, $\vec{\beta}$, and 
$\vec{\alpha}+\vec{\beta}$ of $g$, it can be easily checked that 
\be
m_{\vec{\alpha}+\vec{\beta}}^2 =  m_{\vec{\alpha}}^2 +  
m_{\vec{\beta}}^2  + 2\> m_{\vec{\alpha}} m_{\vec{\beta}} \cosh
\sigma_{\vec{\alpha},\vec{\beta}}\>,
\ee
where
\be
\sigma_{\vec{\alpha},\vec{\beta}} = {1\over2} \ln  {(\vec{\alpha}\cdot
\vec{\lambda}_+) (\vec{\beta}\cdot
\vec{\lambda}_-)\over (\vec{\alpha}\cdot
\vec{\lambda}_-)(\vec{\beta}\cdot \vec{\lambda}_+)}\>.
\ee
These equations, together with~\rf{MassF}, establish the relationship
between the arbitrary constants $\vec{\lambda}_\pm$ and the
different scales that determine the semiclassical mass spectrum of stable and   
unstable particles.

In~\cite{DECAY}, it was conjectured that the exact solution of the simply
laced HSG theories is provided by a diagonal $S$-matrix. In this solution, the
exact spectrum of stable particles coincides with the semi-classical spectrum of
soliton particles of the HSG theory associated with the simple roots of the
algebra, and there is an independent mass scale attached to
each simple root or, equivalently, to each node of the Dynkin diagram
of~$g$:
$m_1,\ldots, m_{r_g}$. Following~\cite{OTBA}, particles will be labelled
by two quantum numbers
$(a,i)$, with $1\leq  a\leq k-1$ and $1\leq i\leq
r_g$, and $S_{ab}^{ij}(\theta)$ will be the two-particle scattering
amplitude corresponding to the process where the particle $(a,i)$
initially is on the left-hand-side of the particle $(b,j)$. For particles
associated to the same simple root, $i=j$, the amplitude is
provided by the minimal $S$-matrix associated to $A_{k-1}$
$$
S_{ab}^{ii}(\theta)= S_{ab}^{A_{k-1}}(\theta)= (a+b)_\theta \>
(|a-b|)_\theta $$
\be
\times\prod_{n=1}^{{\rm min\/}(a,b)-1}
(a+b-2n)^2_\theta
\>,
\lab{Min}
\ee
where we have introduced the block notation $(x)_\theta =
\sinh{1\over2}(\theta +i{\pi x\over k})/
\sinh{1\over2}(\theta -i{\pi x\over k})$.
On the other hand, the scattering between solitons associated to different
simple roots is described by
$$
S_{ab}^{ij}(\theta)= (\eta_{i,j})^{ab}
\prod_{n=0}^{{\rm min\/}(a,b)-1} (-|a-b|-1-2n)_{\theta+\sigma_{ij}}\>,
$$
\be
\not= S_{ba}^{ji}(\theta) \hskip4truecm
\lab{New}
\ee
if $\vec{\alpha}_i + \vec{\alpha}_j$ is a root of $g$, and by
$S_{ab}^{ij}(\theta) =1$ otherwise. In this equation
$\sigma_{ij} = -\sigma_{ji}$ $(\sigma_{ii}=0)$ are $r_g-1$ free real
parameters attached to the links of the Dynkin diagram of $g$. They
determine the position of the resonance poles or, equivalently, the mass
scales of the spectrum of unstable particles. In particular,
$S_{aa}^{ij}(\theta)$ has a resonance pole at $\theta= \sigma_{ji}
-i\pi/k$ which, for
$\sigma_{ji}>0$, should correspond to the unstable soliton particle
with mass $M_{\vec{\alpha}_i +\vec{\alpha}_j}[a]$ in the semiclassical
$k\gg h_g$ limit. The existence of the
$2r_g -1$ free parameters $m_i$ and $\sigma_{ij}$ is a consequence of the
freedom to choose $\Lambda_\pm$ in the classical action~\rf{Act}. Finally,
$\eta_{i,j}= \eta_{j,i}^\ast$ for $i\not=j$ ($\eta_{ii}=1$) are arbitrary
$k$-th roots of~$-1$ whose presence is required to satisfy both the crossing
relations and the bootstrap equations~\cite{DECAY}. 

Eqs.~\rf{Min} and~\rf{New} admit the
following integral representation~\cite{OTBA}
$$
S_{ab}^{ij}(\theta) = \sqrt{\eta_{i,j}^{-2k\> {\cal C}_{ab}^{-1}}}\> \exp
\int {dt\over t}\> {\rm e}^{-it(\theta+ \sigma_{ij})}\qquad
$$
\be
\times \left(2\cosh {\pi t\over k}  - {\cal I}^{(g)}\right)_{ij}  
\left(2\cosh {\pi t\over k} - {\cal I}\right)^{-1}_{ab} \>, 
\lab{IntS}
\ee
where ${\cal C}= 2-{\cal I}$ and ${\cal C}^{(g)} = 2-{\cal I}^{(g)}$ are
the Cartan matrices of
$A_{k-1}$ and
$g$, respectively. The scattering matrices of the simply laced HSG theories
have been recently generalized in a Lie algebraic sense in~\cite{CVS},
where a new factorised
$S$-matrix theory is associated to each pair $(\widetilde{g}| g)$ of
simply laced Lie algebras. In this construction, the $G_k$--HSG $S$-matrix
corresponds to the pair $(A_{k-1}| g)$. 

So far, the relationship between the proposed $S$-matrices and the simply
laced HSG theories has been checked by means of the Thermodynamic Bethe
ansatz (TBA)~\cite{OTBA} and, more recently, through the explicit
calculation of the Form Factors~\cite{OFF}. Assuming
that all particles are of fermionic type ($S_{aa}^{ii}(0)=-1$), the TBA
equations read
\be
\epsilon_a^i(\theta)+ \sum_{b,j} \Phi_{ab}^{ij}\ast
L_b^j (\theta) = R\> M_a^i  \cosh\theta\>, 
\lab{TBAE}
\ee
where $M_a^i= m_i \sinh (\pi a/k)$ and $L_b^j(\theta) = \ln \bigl(1
+ \exp -\epsilon_b^j(\theta)\bigr)$. Recall that the scattering
amplitudes given by eqs.~\rf{Min}--\rf{IntS} are not parity symmetric.
Actually, it is easy to check that they satisfy the Hermitian
analyticity condition
$S_{ab}^{ij}(\theta) = \bigl[S_{ba}^{ji}(-\theta^\ast)\bigr]^\ast \not=
S_{ab}^{ij}(\theta)$ which, for real values of the rapidity, implies that
\be
\Phi_{ab}^{ij}(\theta) = -i\>{d\over d\theta} \ln
S_{ab}^{ij}(\theta) =  \Phi_{ba}^{ji}(-\theta) \not=
\Phi_{ab}^{ij}(-\theta)
\ee
and, hence, $\epsilon_a^i(\theta)\not= \epsilon_a^i(-\theta)$.
Consequently, the finite size scaling function is given by
\be
c(R) ={3 R\over \pi^2} \sum_{a,i} M_a^i\int_0^\infty d\theta\>
\cosh\theta\> \Bigl( L_a^i(\theta)  + L_a^i(-\theta)\Bigr)\>. 
\ee

An important quantity that is provided by the TBA analysis is the value of
$c(R)$ in the $R\rightarrow0$ limit, which corresponds to the the central charge
of the CFT that governs the ultra-violet (UV) limit of the
$S$-matrix theory, assuming that it is unitary. Taking into
account that the kernels $\Phi_{ab}^{ij}(\theta-\sigma_{ij})$ are strongly
peaked at $\theta=0$, one can prove that its value is given by~\cite{OTBA} 
\be
\lim_{R\rightarrow0} c(R) = {6\over \pi^2}\> \sum_{a,i} {\cal L\/}(f_a^i)
\lab{Limit}
\ee
where ${\cal L\/}(x)$ is Roger's dilogarithm function, and $f_a^i =
\bigl(1+\exp \epsilon_a^i(0)\bigr)^{-1}$, for $a= 1,\ldots, k-1$, and $i=
1,\ldots, r_g$, satisfy the constant TBA equations
\be
\sum_{b=1}^{k-1} {\cal C}_{ab} \ln f_b^i = \sum_{j=1}^{r_g}  {\cal
C}_{ij}^{(g)} \ln (1-f_a^j)\>.
\ee
These equations were considered before in the context of the `restricted
solid-on-solid' models~\cite{RSOS} and solved in terms of Weyl
characters, with the result that the value of $c(R)$ for $R\rightarrow0$
indeed coincides with the central charge of the CFT associated with the
coset $G_k/U(1)^{r_g}$ for any $G$ and $k$, which is given in eq.~\rf{CFT}. 

All this confirms that the $S$-matrices
defined by eqs.~\rf{Min}--\rf{IntS} describe massive integrable
perturbations of the theory $G$-parafermions at level-$k$ for any (finite)
value of the $2\> r_g -1$ free parameters $(m_i, \sigma_{jl})$, which
strongly supports the conjectured connection with the HSG models. 
Additional support is provided by the recent application of the Form Factor
program to the $SU(3)_2$--HSG theory in~\cite{OFF}, which allows the
authors to calculate not only the value of the UV central charge $(6/5)$
using the Zamolodchikov's $c$-theorem, but also the conformal dimension of
various operators including the dimension of the perturbing operator
$(3/5)$. 

Nevertheless, this is not the only information that can be obtained from
the TBA analysis. For finite values of $R$, the finite size scaling
function $c(R)$ shows a rather characteristic `staircase pattern' where
the number of steps is finite and their position is fixed by the value
of the different free parameters entering the definition of the $S$-matrix.
This suggests the interpretation of the staircase pattern as a consequence
of the change in the number of effective light degrees of freedom
produced by the decoupling of the heavy particles, both stable and
unstable, compared to the the scale given by the temperature $T\simeq 1/R$.
This provides a strong argument in favour of the interpretation of the
resonance poles as a trace of the presence of unstable particles in the
spectrum, in agreement again with the expected properties of the HSG
models. It is worth noticing that the original staircase
models~\cite{StairS} do not allow such a direct physical interpretation
for the observed staircase pattern.

Although $c(R)$ has to be calculated numerically, it is
possible to understand analytically the physical origin of the staircase
behaviour. It will be convenient to separate the TBA kernel in two
different parts:
\be
\Phi_{ab}^{ij}(\theta)  = \cases{\phi_{ab}(\theta)&; $i=j$, \cr
\noalign{\vskip 0.2truecm}
{\cal I}^{(g)}_{ij}\> \psi_{ab}(\theta+\sigma_{ij}) &;
$i\not=j$,\cr}
\ee
where, according to~\rf{Min}, $\phi_{ab}(\theta)=\phi_{ab}^{A_{k-1}}(\theta)$
is just the TBA kernel of the minimal $S$-matrix associated to $A_{k-1}$. Then,
the TBA equations~\rf{TBAE} become
$$
\epsilon_a^i(\theta) + \sum_{b=1}^{k-1}\Bigl[ \phi_{ab}\ast
L_{b}^i(\theta)  + \sum_{j\not=i} {\cal I}^{(g)}_{ij}\> \psi_{ab}\ast
L_{b}^j(\theta+ \sigma_{ij}) \Bigr] 
$$
\be
= RM_{a}^i \cosh\theta\>,\hskip 2truecm
\lab{TBAPlus}
\ee
which shows that whenever the term involving the
kernel $\psi_{ab}$ is negligible we are left with the TBA equation
for the minimal $A_{k-1}$ $S$-matrix theory: $L_{a}^i(\theta) \simeq L_a^{\rm
min}(\theta)$. Moreover, standard arguments show that
$L_{a}^i(\theta) \simeq 0$ if $m_i\gg 2/R$ due to the dominance
of the energy term, which allows one to show that the effect of $\psi_{ab}$ is
negligible for 
\be
|\sigma_{ij}|\gg -\ln(R^2 m_i m_j/4)\>.
\ee 

Recall that
$S_{ab}^{ij}(\theta)$ exhibits complex poles whose position is of the form
$\theta = \sigma_{ji} - i\pi n/k$, where $n$ is some positive integer. For
$\sigma_{ji}\gg0$  and $n\ll k$, corresponding to $\Gamma\ll M$
in~\rf{BW}, they are conjectured to indicate the
presence of unstable particles whose mass scale is given by
\bea
m_{R_{ij}}^2 &=& m_i^2 + m_j^2 + 2 m_i m_j \cosh |\sigma_{ij}| \nn
&\simeq& m_i m_j {\rm e\>}^{|\sigma_{ij}|}\>.
\ena
Thus, the condition to neglect the term involving $\psi_{ab}$
in~\rf{TBAPlus} is just $m_{R_{ij}}  \gg 2/ R$
which means that the unstable particles related to
$|\sigma_{ij}|$ are heavy and,
hence, they are decoupled from the effective theory at the scale fixed by
$R$. Notice that the previously mentioned result that $L_{a}^i(\theta)
\simeq 0$ if $m_i\gg 2/R$ admits a similar interpretation by changing
unstable particles by stable ones. Taking into account all this, together
with eq.~\rf{Limit}, we can sketch the following conjecture about
the staircase behaviour of the scaling function:
\begin{itemize}
\item[1)] $c(R) \simeq0\quad$ if
\item[] ${2\over R}\ll m_{i_1},
\ldots, m_{i_{r_g}}, m_{R_{ij}}$\quad (deep IR)
\item[2)] 
$c(R) \simeq n\> {2(k-1)\over k+2}\quad$ if

$m_{i_1},\ldots,m_{i_n}\leq {2\over R}\ll m_{i_{n+1}},\ldots,
m_{i_{r_g}}, m_{R_{ij}}$, 
\item[] for $n=1,\ldots, r_g$,
\item[3)]  $c(R) \simeq {k-1\over k+h_g}\> h_g r_g\quad$ if

$m_{i_1},
\ldots, m_{i_{r_g}}, m_{R_{ij}} \ll {2\over R}$\quad (deep UV), 

and always for all $i\not= j$. 
\end{itemize}
Nevertheless, notice that these qualitative arguments do not explain the
behaviour of $c(R)$  in the wole range
$0\leq R<\infty$ and, hence, do not allow one to predict the
precise number of steps.

In~\cite{OTBA}, the scaling function $c(R)$ was numerically calculated for
the $SU(3)_k$--HSG models with $k=2,3,4$ and the results are in agreement
with our conjecture. Since
$r_g=2$, there are only three free parameters: $m_1$, $m_2$,
and $\sigma_{21}=-\sigma_{12}\equiv \sigma$, which means that there is
only one mass scale for the unstable particles. The maximum
number of steps turns out to be three in this case, which corresponds to the
situation when
$m_1\ll m_2\ll m_{R_{12}}$. In contrast, if $m_1\simeq m_2$ and
$\sigma\simeq0$ there are no steps at all. A remarkable result is that,
whenever the unstable particles are much heavier than the stable ones,
$m_{R_{12}} \gg m_1, m_2$, the UV limit of the
$SU(3)_k$--HSG models may be viewed alternatively as a massless IR
$\rightarrow$ UV flow between two different coset conformal field theories:
\be
{SU(2)_k\over U(1)} \times {SU(2)_k\over U(1)} \longrightarrow
{SU(3)_k\over U(1)^2}\>.
\ee
As a particular case, the flow between the tricritical Ising and
the critical Ising model is recovered as a subsystem for $k=2$.

In conclusion, the results obtained so far from the TBA
analysis~\cite{OTBA} and the calculation of Form Factors~\cite{OFF} 
confirm the conjectured relationship between the $S$-matrix theories proposed
in~\cite{DECAY} and the simply laced HSG theories. Further support could be
obtained by extending the explicit calculations to other Lie groups
different from
$SU(3)$. Moreover, it would be extremely interesting to investigate
directly the connection of the resonance poles with the unstable particles
of the HSG theories.

\acknowledgments
This research is supported
partially by CICYT (AEN99-0589), DGICYT (PB96-0960), and the EC Commission
(TMR Grant FMRX-CT96-0012).

\end{document}